\documentclass[11pt,reprint,amsmath,amssymb,aps]{revtex4-2}

\usepackage{graphicx}
\usepackage{dcolumn}
\usepackage{bm}
\usepackage{braket}
\usepackage{xcolor}
\usepackage{tablefootnote}
\usepackage{array}
\usepackage{setspace}
\newcolumntype{C}[1]{>{\centering\arraybackslash}p{#1}}

\newcommand{\red}{\color{red}}
\def\red#1 {\textcolor{red} {#1} }
\newcommand{\br}{\langle}
\newcommand{\ke}{\rangle}
\newcommand{\vare}{\varepsilon}
\newcommand{\beb}{\bar{\beta}}
\newcommand{\bebb}{\bar{\bar{\beta}}}
\newcommand{\Wpol}{W_{\mathrm{pol}}}
\newcommand{\vwpol}{v_{W\,\mathrm{pol}}}

\begin{document}

\title{Optimized Attenuated Interaction: Enabling Stochastic Bethe-Salpeter Spectra for Large Systems}

\author{Nadine C Bradbury}
 \email{nadinebradbury@ucla.edu}
 \affiliation{Department of Chemistry and Biochemistry, UCLA, Los Angeles CA 90095-1569 USA}

\author{Tucker Allen}
 \affiliation{Department of Chemistry and Biochemistry, UCLA, Los Angeles CA 90095-1569 USA}

\author{Minh Nguyen}
 \affiliation{Department of Chemistry and Biochemistry, UCLA, Los Angeles CA 90095-1569 USA}

\author{Khaled Z Ibrahim}
\affiliation{Computer Science Department, Lawrence Berkeley National Laboratory,
One Cyclotron road, Berkeley, CA 94720, USA}

\author{Daniel Neuhauser}
 \affiliation{Department of Chemistry and Biochemistry, and California Nanoscience Institute, UCLA, Los Angeles CA 90095-1569 USA}

\date{\today}

\begin{abstract}
We  develop an improved stochastic formalism for the Bethe-Salpeter equation, based on an exact separation of the effective-interaction $W$ to two parts, $W=(W-v_W)+v_W$ where the latter is formally any translationally-invariant interaction $v_W(r-r')$. When optimizing the fit of  $v_W$ exchange kernel to $W$, by using a stochastic sampling of  $W$, the difference $W-v_W$ becomes quite small.  Then, in the main BSE routine, this small difference is stochastically sampled.  The number of stochastic samples needed for an accurate spectrum is then largely independent of system size. While the method is formally cubic in scaling, the scaling prefactor is small due to the constant number of stochastic orbitals needed for sampling $W$.
\end{abstract}

\maketitle

\section{Introduction}
\vspace{-10pt}

The Bethe-Salpeter Equation (BSE), a many-body perturbation theory method, is becoming increasingly popular for predicting optical spectra of chemical systems.~\cite{Blase2020} Physically, BSE goes beyond time-dependent density functional theory (TDDFT) by the inclusion of the correct long range exchange kernel in the effective interaction $W$. Numerically, however, the BSE is quite expensive,
mostly due to the cost of generating  the two-electron integrals of the effective interaction $W$, which scales  formally as  $\mathrm{O}(N^4)$, or in specially optimized cases $\mathrm{O}(N^3)$,\cite{Ljungberg2015,Duchemin2019}, where $N$ is the number of electrons. Due to the steep scaling, the BSE is  typically applied for systems with up to about 100 valence and conduction states. However, thanks to many advancements in the algorithms used,\cite{Rocca2012-WEST,BerkeleyGW,YAMBObse,Frster2021} the method was recently applied to a system of nearly 2000 total electrons.~\cite{Frster2022}

Recently we developed a numerically efficient approach to the BSE that relies on a stochastic evaluation of $W$.~\cite{Bradbury2022} Adopting the usual Tamm-Dancoff approximation (TDA), $W$ is then applied on all pairs of occupied-occupied states (see later for details), and since its evaluation is linear in system size,  systems with hundreds of active electrons become feasible.

In this work we go a step beyond, and show that not only is the action of $W$  obtained efficiently with a stochastic approach, but, equally important, the explicit matrix elements can be replaced by a stochastic sampling of the sea of occupied-occupied pairs.  This, in principle, limits the major cost of the BSE to quadratic scaling, thereby opening the possibility of very large systems.

A key in our proposed approach is the numerically exact rewriting of the action of $W$ by subtracting and adding a simple Coulomb-like interaction $v_W$.  Thus, the stochastic sampling only needs to be applied on this small difference $W-v_W$, with the bulk of the action of $W$ done by $v_W$. This stabilizes the stochastic approach ensuring that for larger systems we do not need more stochastic samples to represent $W$.

Choosing an analytical Coulombic-like interaction to substitute for $W$ has been done before in some efficient implementations of the BSE,\cite{Fuchs2008,Rabani2015} but here we choose an optimized $v_W$ which is fitted to the actual $W$ of each system. The use of $v_W$ is also reminiscent of TDDFT based approaches with long range exchange and a polarizable medium that mimics the dielectric function.~\cite{Begam2020} In our work, since $v_W$ is built from $W$, the ab initio nature of the BSE is retained, while still reducing the complexity of the exchange to be similar to traditional Fock exchange.

Using $v_W$ by itself also gives fairly reasonable spectral results.  Thus, our work is not only a numerically more efficient way to calculate the BSE spectra, but gives an alternative, fairly cheap algorithm, at the same cost as time-dependent Hartree-Fock (TDHF), which itself can be done cheaply with a stochastic approach,\cite{Neuhauser2015sX, Gao2015,Rabani2015} that has an improving accuracy for increasingly large systems. 

The paper is organized as follows. 
The methodology is reviewed in Section \ref{sec:method}. Section \ref{sec:results} shows results for a variety of medium to large carbon based systems.  Conclusions follow in Section \ref{sec:con}.

\section{Methods\label{sec:method}}
\subsection{Iterative BSE formulation\label{sec:bse}}

We first overview the methodology for obtaining spectra from the BSE for a given $W$. 

The starting point is a closed shell system with $2 N_{\rm occ}$ electrons. The exciton (electron-hole) basis is a set of $n_o$ occupied (valence) states $\phi_i, \phi_j,...,$ times a set of $n_c$ conduction states, $\phi_a, \phi_b,...,$ which are eigenstates of a zero order (typically DFT) Hamiltonian.  Further, we use the TDA, although the approach is generalizable to the full BSE.

The starting optically excited vector $f^0$ is an overlap of the excitons with the coordinate in the direction of the laser polarization (labeled here as $\hat{x}$ for simplicity):
\begin{equation}
    f^0_{ja} = \br \phi_a | x | \phi_j \ke.
\end{equation}

The spectrum is then obtained from a matrix element of the frequency resolved Liouvilian operator, $A$, governing the motion of the excitons
\begin{equation}\label{eq:abs}
    \sigma(\omega) \propto 
    \omega \br f^0 | \delta (A - \omega )   | f^0 \ke,
\end{equation}
where the broadened delta function is obtained by a Chebyshev 
series, 
\begin{equation} \label{eq:cheby}
     \delta (A - \omega )|f^0 \ke = \sum_n c_n(\omega) |f^n \ke,
\end{equation}
where $c_n$ are numerical coefficients and $f^n$ are Chebyshev vectors, obtained by iteratively applying $A$ on $f^0$.  In practice we find that the best results are obtained by simple smoothly-decaying weights, in the spirit of those used in cite\cite{Weisse2006}
\begin{equation}
c_n(\omega) = \left|\frac{d\theta_\omega}{\omega}\right|\mathrm{cos}^2\left(\frac{\pi n}{2N_{\mathrm{cheby}}}\right)  \mathrm{cos}(n \theta_\omega),
\label{eq:c_n_simple}
\end{equation}
where $N_{\mathrm{cheby}}$  is the number of Chebyshev terms used, which determines the frequency resolution.  Here we introduced the Chebyshev angle $\theta_\omega \equiv \mathrm{cos}^{-1}({\omega}/{\delta A})$, while $\delta A$  is an upper bound on the half-width of the spectrum of $A$.  Note that without the $\left|{d\theta_\omega}/{\omega}\right|$ term, these weights would yield a delta function in $\theta_\omega$, and this term converts the overall function to a delta function over $\omega$.

Formally $A$ is made from three terms: diagonal, Hartree and the so-called direct term (in a somewhat confusing notation, since it resembles Fock exchange):
\begin{equation}
    A_{ia,jb}= (\vare_a - \vare_i + \Delta ) \delta_{ij} \delta_{jb} + \kappa(ia|jb) - (\phi_a \phi_b | W|\phi_i \phi_j ),
\end{equation}
where we introduced the electron and hole energies associated with the respective zero order eigenstates, while the round brackets refer to an $(r,r|...|r',r')$ notation. $\Delta$ is a scissors shift that corrects the gap to match accurate $\mathrm{GW}$ calculations, and could, if wished, depend on the exciton ($i,a$) indices -- as is especially important for small systems.~\cite{Gui2018,McKeon2022}  In practice we use the cheap sGW, i.e., stochastic $\mathrm{GW}$ (see below) to calculate the scissors term,\cite{Neuhauser2014sGW,Vlek20181} and for further accuracy we implement the scissor-shift self-consistent $\mathrm{GW_0}$ approach, labeled $\Delta\mathrm{GW}_0$,\cite{Vlek20183} which post-processes the results of sGW and generally raises the gap by a few tenths of eV.

The Hartree integral is (assuming real orbitals):
\begin{equation}
    (ia|jb)= \int \phi_i(r)\phi_a(r) v(r-r') \phi_j(r')\phi_b(r') dr dr',
\end{equation}
where $v(r-r')=1/|r-r'|$ is the Coulomb interaction, while  $\kappa$ is 2 for singlet excitations, and 0 for triplet excitations. 
Finally, the most numerically costly part involves the effective interaction 
\begin{equation}
    (\phi_a \phi_b | W|\phi_i \phi_j )= \int \phi_a^*(r)\phi_b^*(r) W(r,r') \phi_i(r')\phi_j(r') dr dr'.
\end{equation}
Note that $W$ refers to the static part of the effective interaction, and we ignore here the effects of the dynamic part.

Numerically, one acts with $A$ on an arbitrary vector $f$ as follows:
\begin{equation}
\label{eq:Afij}
\begin{split}
g_{ia} & \equiv 
\left( A f \right)_{ia} = \\ 
  & (\vare_a - \vare_i + \Delta )f_{ia} 
   + \frac{\kappa}{2}\br \phi_a |\delta v_H | f_i \ke 
   - \br \phi_a | y_i \ke,
   \end{split}
\end{equation}
where the grid-representation of the exciton is
\begin{equation}
    f_i(r) = \sum_b f_{ib} \phi_b(r)
\end{equation}
while the exciton Coulomb density, $\delta n(r) = 4 \sum_j f_j(r) \phi_j(r)$, is used to generate the Hartree potential,
\begin{equation}
    \delta v_H(r) = \int  \frac{\delta n(r')}{r-r'} dr'.
\end{equation}

The numerically expensive part in Eq. (\ref{eq:Afij}) comes from the direct term, involving the action of the effective interaction,
\begin{equation} \label{eq:bse_det}
    y_i(r) \equiv \sum_j W_{ij}(r) f_j(r),
\end{equation}
where
\begin{equation}
     W_{ij}(r) \equiv
    \int W(r,r') 
    \phi_i(r')\phi_j(r') dr'.
\end{equation}

In our recent work,\cite{Bradbury2022} we used the stochastic time-dependent Hartree (i.e., stochastic $W$) approach,\cite{Gao2015} developed originally for sGW,\cite{Neuhauser2014sGW,Vlek20181} to evaluate each specific $W_{ij}$ function in linear scaling; see Ref. (\cite{Bradbury2022}) for full details on this step of the method. The application of stochastic $W$ makes is feasible to study systems with up to several hundred valence states. Nevertheless, as there are $\simeq N_v^2/2$ such terms for $N_v$ valence states, the overall cost is cubic in system size with a large pre-factor, so that including more than $\approx 300$ valence states will be numerically challenging.

\subsection{Stochastic evaluation of matrix elements}
To overcome the scaling problem, we use a stochastic representation of the sum.  Specifically, we define a stochastic process, made from ``instances".  For each such instance, we define two independent stochastic vectors,
\begin{equation}
\label{eq:beb}
\begin{split}
    \beb(r) = \sum_l \beb_l \phi_l(r), & \\ 
    \bebb(r) = \sum_l \bebb_l \phi_l(r), &
    \end{split} 
\end{equation}
where $\beb_l = \pm 1$, $\bebb_l = \pm 1$.

Using \begin{equation}
\Big\{ \beb_i \beb_j \Big\} = \Big\{\bebb_i \bebb_j \Big\}= \delta_{ij},
\end{equation}
where curly brackets denote an average over many stochastic instances, it follows that 
\begin{equation}
  \Big\{ \beb_i \bebb_j \beta(r)
  \Big\}
  = \phi_i(r) \phi_j(r),
\end{equation}
where
\begin{equation}
    \label{eq:beta}
    \beta(r)\equiv \beb(r) \bebb(r).
\end{equation}

Inserting the relations above to 
the numerically expensive effective potential term in Eq. (\ref{eq:Afij}), the latter becomes
\begin{equation}
\label{eq:u_Wbetaf}
   y_i(r) =  \Big\{ {\beb}_i 
   \br r | W |\beta \ke 
    f_{\bebb}(r) 
    \Big\},
\end{equation}
where we defined
\begin{equation}
    f_{\bebb}(r) = \sum_j \bebb_j f_j(r),
\end{equation}
while
$
    \br r | W |\beb \bebb \ke \equiv \int W(r,r')\beb(r) \bebb(r) dr'.
$

The resulting algorithm is thus quite simple.  A large but finite number of stochastic instances, $N_\beta$, is defined.  Then, using one applies  $W$ (calculated itself stochastically) on the stochastic representation of the valence density  $\beb(r) \bebb(r)$, to yield a set of $N_\beta$ vectors, $\br r | W |\beb \bebb \ke$, which is stored and used in the Chebyshev iterative step, $f\to Af$.  The formulae are further detailed in the next section.

\subsection{Optimized attenuated interaction\label{sec:vw}}
\subsubsection{Sampling a small difference}
The formalism above is clearly a member of our stochastic approaches to quantum chemistry.~\cite{Baer2022-review, Neuhauser2014sDFT,Neuhauser2014sGW, Neuhauser2015sX,Rabani2015} The key in these approaches is the replacement of individual molecular orbitals by random orbitals, that are stochastic combination of individual orbitals.  For example, a valence orbital is replaced by a stochastic combination of valence orbitals, etc.

A key practical point in this paradigm is that it is best to stochastically sample numerically small quantities. This is best achieved by sampling just the difference between the desired quantity and a simpler one, i.e., writing
\begin{equation}
    W= \{W- v_W\} + v_W
\end{equation}
where curly brackets indicate again a statistical average and $v_W(r,r')$ is an interaction which is ``cheap" to act with.  Here we use the simplest such form, a translationally invariant two-body interaction,
\begin{equation}
    v_W(r,r')=v_W(r-r')
\end{equation}
The specifics of $v_W$ are delineated later.

Using this decomposition, the action of $W$, Eq. (\ref{eq:u_Wbetaf}), is modified to
\begin{equation}
\label{eq:u_Wv_betaf}
   \begin{split}
   y_i(r) =  \Big\{ {\beb}_i \br r | W-v_W |\beta \ke 
    f_{\bebb}(r) \Big\}   
    + \sum_{j} f_j(r) \br r | v_W | \phi_i \phi_j \ke & \\
    = \Big\{ {\beb}_i \br r | W- v_W |\beta \ke  
    f_{\bebb}(r) \Big\} 
    + \sum_{j} f_j(r) v_{W,ij}(r), &
    \end{split}
\end{equation}
where
$v_{W,ij}(r)=\int v_W(r,r') \phi_i(r') \phi_j(r') dr'$.

\begin{figure}[ht]
    \centering
\includegraphics{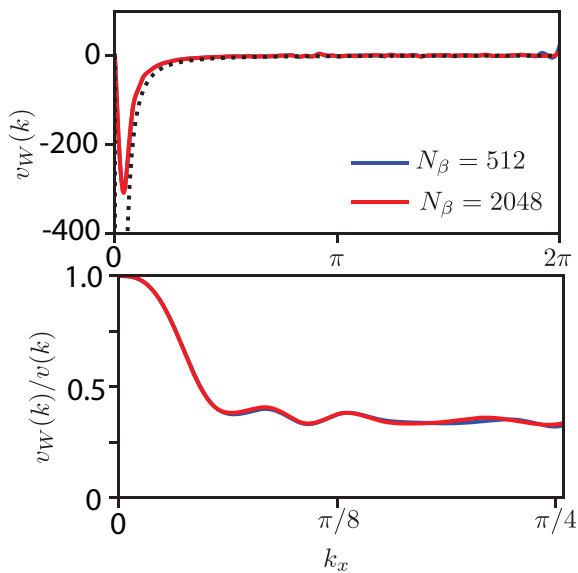}
    \caption{(Top) The fitted $\vwpol(k)$ potentials (red) and the bare Coulomb interaction $v(k)$ (black dots) for $\mathrm{C_{96}H_{24}}$, shown for a range  of $k_x$, for $k_y=k_z=0$.
    (Bottom) The ratio $v_W(k)/v(k)$ for this system, calculated for the same $k$-values range as in the top panel.  The results converge quickly with the number of stochastic sampling functions, $N_\beta$.}
    \label{fig:vwk}
\end{figure}

\begin{figure*}[ht]
    \centering
    \includegraphics[width=\textwidth]{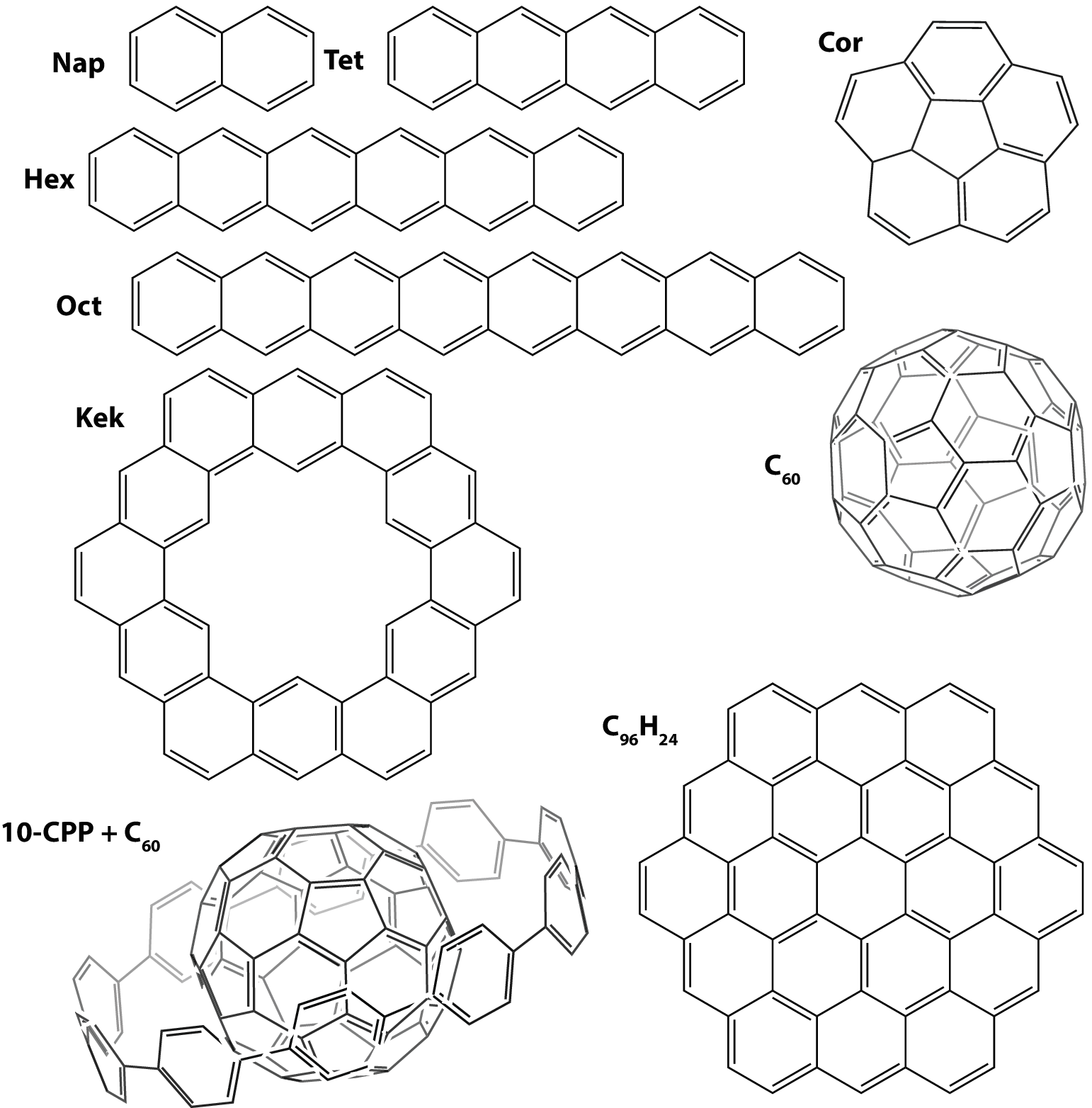}
    \caption{Structures and abbreviations for all the systems used in this paper.}
    \label{fig:sys}
\end{figure*}

\subsubsection{Optimizing the effective interaction potential}
The equation above is exact no matter what $v_W$ is -- a better choice of $v_W$ would simply lead to faster  convergence of the sampling of $\{W-v_W\}$.  
Further, to avoid the singularities of the Coulomb potential we fit only the polarization part, i.e., $W-v_W = \Wpol-\vwpol$, where $\Wpol=W- v(k)$, and similarly for $\vwpol$.   Here $v(k)$ is the Coulomb potential for finite systems, which is obtained with the Martyna-Tuckerman approach;\cite{Martyna1999} this potential is the usual $4 \pi/k^2$  at high momenta but levels off to a finite large value at $k=0$.

Given an arbitrary large system, we can ask what will be the optimized $v_W(r-r')$.  Interestingly, the stochastic paradigm answers that question easily.  Specifically, optimize the functional
\begin{equation}
    J= \sum_{ij}(\phi_i \phi_j |(W-v_W)^2 |\phi_i \phi_j)
\end{equation}
where again $i,j$ are occupied states.
Calculate the sum then stochastically
\begin{equation}
    J= \Big\{ \br \beta | (W-v_W)^2 |\beta \ke \Big\}=
    \Big\{ \int |\br k | W-v_W|\beta \ke|^2 dk \Big\},
\end{equation}
where, as before, $\beta$ is a stochastic combination of the occupied two-electron product terms from Eq. (\ref{eq:beta}).

\begin{figure*}[ht]
    \centering
    \includegraphics{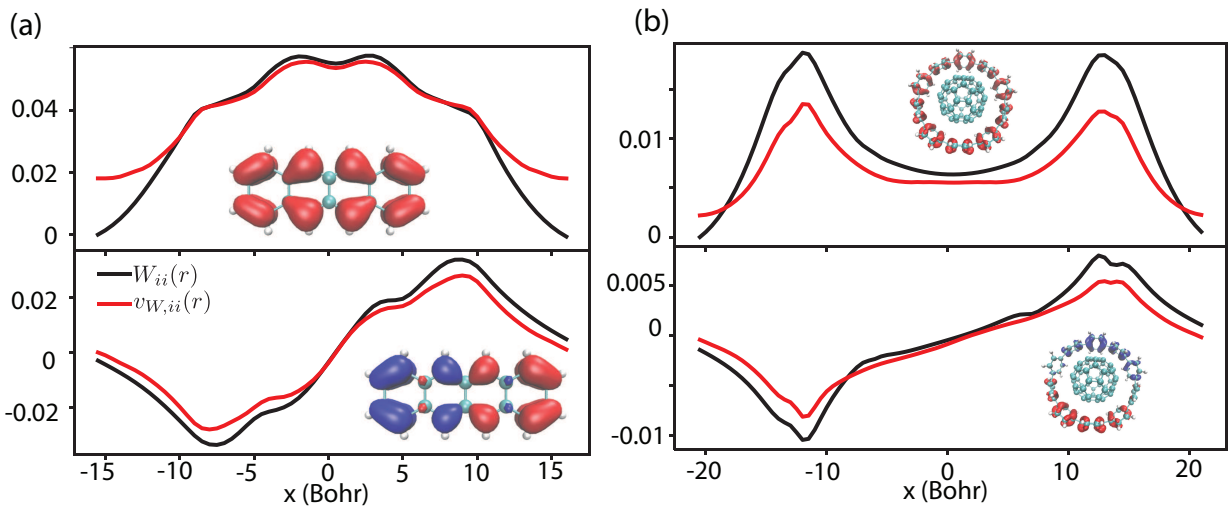}
    \caption{X-axis slice of $\langle r | W|\phi_i \phi_j \rangle$ and $\langle r | v_W|\phi_i \phi_j \rangle$, i.e., the action of the true $W$ (black) and  the optimized $v_W$ (red) on a two-orbital pair density. In the top row $W$ and $v_W$ act on the HOMO density ($i=j=\mathrm{HOMO}$). In the bottom row they act on the pair density of the $\mathrm{HOMO\times HOMO-1}$ orbitals. The left part, Column (a), is for tetracene, while Column (b) shows the same plots for the much larger 10-CPP+$\mathrm{C_{60}}$.}
    \label{fig:W_plot}
\end{figure*}

Since our choice of $v_W$ is diagonal in momentum space,  $\br k | v_W |\beta \ke=v_W(k)\beta(k)$,  
it is easy to show that the optimised $J$, giving $\delta J / \delta v_W^*(k) =0$, is obtained with
\begin{equation}
   \label{eq:u_beta_W_beta}
   v_W(k) =
   \frac{\Big\{ \beta^*(k)\br k |W|\beta\ke \Big\} }
   {\Big\{  \big|\br k |W|\beta \ke \big|^2 \Big\} }.
\end{equation}
In practice a very small numbers of terms, typically $N_\beta\approx 500$, is sufficient to  converge the values of $v_W(k)$. This convergence is demonstrated in Fig. \ref{fig:vwk}.

\subsubsection{Replacing $W$ by the optimized attenuated interaction}

If $v_W$ is a good enough approximation to $W$, so the objective $J$ is sufficiently small, we may even, as mentioned, throw out the stochastic $\{W-v_W\}$ term in Eq. (\ref{eq:u_Wv_betaf}), i.e., approximate 
\begin{equation}
\label{eq:u_Wv_betafa}
   y_i(r) \simeq  
   \sum_j f_j(r) v_{W,ij}(r).
\end{equation}

More generally, we can approximate the full BSE by replacing $W$ by $v_W$, converting thereby the equation to TDHF-like with a modified Fock kernel, where $|r-r'|^{-1}$ is replaced by $v_W$.  Note that simplified forms have been used to approximate $W$, see 
 e.g.,\cite{Fuchs2008}, but here the optimized attenuated interaction is based on the true system-dependent $W(r,r')$, yielding a fully ab-initio approach. We label the resulting method as  Time-Dependent Optimized-Attenuated-Interaction (TDOAI). 

\subsection{Overall Algorithm}\label{sec:alg}
The overall algorithm is then:
\begin{itemize}
    \item First, a set of stochastic-GW calculations on the HOMO and LUMO is performed to find the necessary scissors shift.
    \item Second, a set of $N_\beta$ random representations $\beta$ of the occupied-states product is calculated and stored per Eqs. (\ref{eq:beb}) and (\ref{eq:beta}).
    \item Each of these $\beta$'s is then used as input for a stochastic-GW calculation, yielding the action of the static effective interaction $\br r | W | \beta \ke$. 
    \item The Fourier-components of the optimized attenuated interaction, $v_W(k)$, are then calculated from Eq. (\ref{eq:u_beta_W_beta}).
\end{itemize}

At that point one has the optimized attenuated interaction, but there are still several possibilities for the dynamics, i.e., how to propagate and solve the BSE and with which terms included. We summarize four such possibilities, and  the Results section below exemplifies the first two, which use the Chebyshev approach (based on Eqs. (\ref{eq:abs}), (\ref{eq:cheby}) and (\ref{eq:Afij})) 
\begin{enumerate}
    \item The BSE kernel, in the Tamm-Dancoff approximation, can be calculated by stochastically sampling the $\{W-v_W\}$ difference, Eq. (\ref{eq:u_Wv_betaf}).
    \item Another direction is to ignore the $W-v_W$ term and act only with the  optimized effective interaction$v_W$, i.e., use Eq. (\ref{eq:u_Wv_betafa}) instead of Eq.  (\ref{eq:u_Wv_betaf}).
    \item One could use the first option, but go past the Tamm-Dancoff approximation, i.e., include off-diagonal terms.  The simplest option, without increasing the numerical effort substantially, would be to use the optimized attenuated interaction in the off-diagonal portion of the BSE. Since the off-diagonal BSE term (i.e., the term that goes beyond the Tamm-Dancoff approximation) is quite small, it should be accurate to replace in it $W$ entirely by $v_W$.  This would reduce the numerical cost substantially compared to the full  cost of applying $\{W-v_W\}$ stochastically in the off-diagonal term, which would have required a different samplings of the action of $W$, this time acting on an occupied-unoccupied pair density.
    
    \item Finally, just like the second option above, we could use only the attenuated interaction while  avoiding the Tamm-Dancoff approximation. This could be done by either extending the exciton vector space to go beyond the TDA, or by replacing the Chebyshev method altogether by a full-fledged TDHF-like study that uses stochastic-exchange \cite{Rabani2015} but would employ here the optimized TDOAI exchange-interaction $v_W$; this direction would be pursued in a latter publication.
\end{enumerate}
\section{Results\label{sec:results}}

\begin{figure*}[ht]
    \centering
    \includegraphics{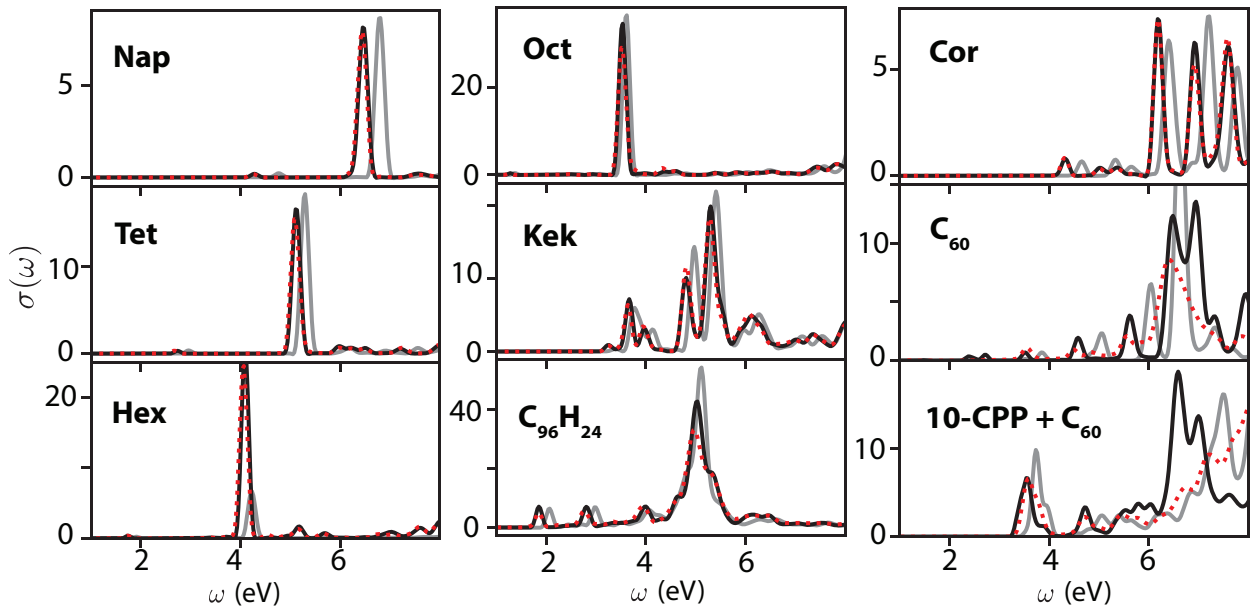}
    \caption{Spectra of singlet excitations for all systems using a BSE with deterministic orbitals (black) and full stochastic $v_W + \{ W-v_W\}$ approach  using $N_\beta = 2000$ (red dots). The TD-OAI calculation, where $v_W$ is used for exchange alone, is shown in grey. For almost all cases, the stochastic approach matches the deterministic optical gap to within 0.02$~$eV or better; the one exception was fullerene, where $N_\beta=5000$ was needed for convergence to 0.08$~$eV, in line with the lower quality of the $v_W$ fit (Table \ref{tab:nv}).}

    \label{fig:spec}
\end{figure*}

\begin{figure}[ht]
    \centering
    \includegraphics{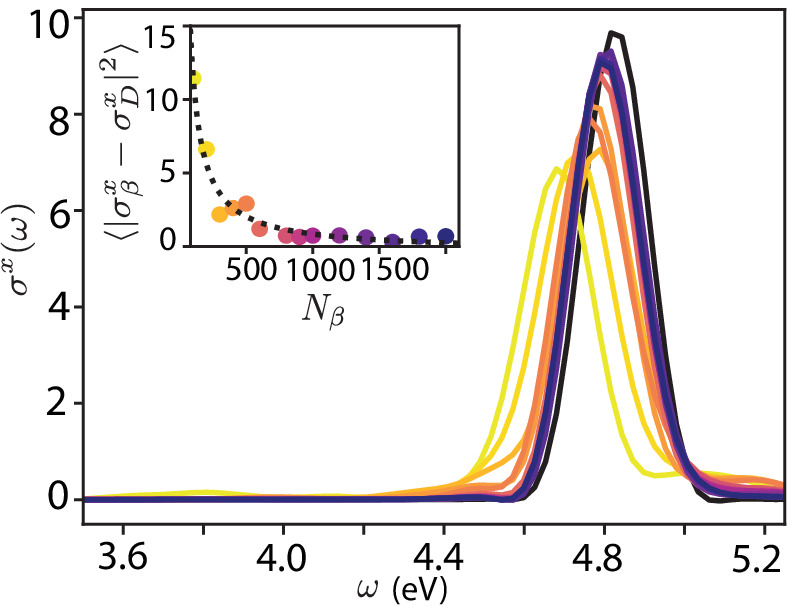}
    \caption{X-polarization spectra for tetracene (zooming in on the dominant spectral peak at 4.75 eV) at varying levels of stochastic approximation (colors), converging to the deterministic BSE (black dashes). The inlay shows the average variance over the 0-6 eV spectral region from the deterministic spectra for each level of approximation, $N_\beta$, with corresponding colors. The dotted curve is a fit to  $1/N_\beta$.}
    \label{fig:conv}
\end{figure}

We demonstrate the new method on a sample set of 
hydrocarbons, including linear acenes, polycyclic-aromatic hydrocarbons (PCH), and fullerene based systems; see Table \ref{tab:nv} and Fig. \ref{fig:sys}. The structures for these molecules were taken from Refs.  \cite{Yang2016,Minameyer2020} and the open source library associated with Ref. \cite{QChem}. For all systems, we use a generous box size extended at least 6 Bohr beyond the edge of the molecule, with a grid spacing of 0.5 Bohr. For the planar molecules, we use a grid size of 15 Bohr in the out-of-plane direction, such that no size effects are seen on the DFT band gap. All DFT calculations were performed with norm-conserving pseudo-potentials and used the PW-MT LDA exchange-correlation functional. \cite{Gao2015,Reis2003, Willand2013}

To determine the correct scissor shift, $\Delta$ in Eq. (\ref{eq:Afij}), we first correct the DFT band gap through a stochastic $\mathrm{GW}$ calculation,\cite{Neuhauser2014sGW,Vlek20181,Vlek20182} done self-consistently.~\cite{Vlek20183} Further, the dielectric correction $W(k\to 0)$, was determined by a linear fit of the BSE spectra at different grid sizes.~\cite{Bradbury2022, Onida1995,Rozzi2006}  The final scissor shift is then the sum of the dielectric correction and the $\mathrm{GW}$ band-gap correction. 

The calculations of the action of $W$ on either deterministic  or stochastic DFT orbital pairs, $\langle r|W|\phi_i\phi_j\rangle$ and $\langle r | W|\beta\rangle$ respectively, were done with only 10 stochastic time-dependent orbitals. Refer to Ref. \cite{Bradbury2022} for explicit details of this step. The sGW calculations were done with a broadening of 0.1 Hartree. A time-step $dt = 0.1$ a.u. was used for a split-operator propagation, and ``cleaning" (i.e., projection of the excited component of the orbitals to be orthogonal to the occupied space -- see \cite{Bradbury2022}) was done every 10 steps.

\begin{table}
\caption{The grid size, number of occupied orbitals, and the chosen number of valence and conduction subset sizes for each system. The final column shows the remaining fraction of the polarization $\Wpol$ interaction not captured by the optimally fitted $\vwpol$ interaction.}
\label{tab:nv}
\centering
\begin{tabular}{|c|cccc|C{2.3cm}|}
\hline
System & $N_g$ & $N_o$ & $N_v$ & $N_c$ &$\frac{\langle(\Wpol-\vwpol)^2\rangle}{\langle \Wpol^2 \rangle}$\\
\hline
Nap                      & 50,688  & 24   & 16   & 40   & 0.18                                                                                                  \\
Tet                      & 76,800  & 42   & 24   & 64   & 0.13                                                                                                  \\
Hex                      & 113,520 & 60   & 36   & 80   & 0.11                                                                                                  \\
Oct                      & 132,000 & 78   & 45   & 100  & 0.10                                                                                                  \\
Cor                      & 69,984  & 45   & 27   & 70   & 0.18                                                                                                  \\
$\mathrm{C_{60}}$        & 195,112 & 120  & 64   & 120  & 0.25                                                                                                  \\
Kek                      & 147,000 & 108  & 54   & 110  & 0.09                                                                                                  \\
$\mathrm{C_{96}H_{24}}$  & 324,480 & 204  & 100  & 500  & 0.09                                                                                                  \\
10-CPP+$\mathrm{C_{60}}$ & 381,024 & 260  & 100  & 500  & 0.13                                                 \\                                     
\hline
\end{tabular}
\end{table}

The number of samples needed for a deterministic calculation is $N_v(N_v+1)/2$ for $N_v$ valence orbitals. In calculations where $W$ is acting on stochastic orbitals, $N_\beta $=2000 was generally used. 

We first discuss the convergence of the fitted  $v_W$  and how does it compare with $W$.

Figure \ref{fig:vwk} shows the convergence of the fitting of the polarization portion of $v_W$.  The top sub figure shows $\vwpol$ in comparison to (minus) the bare Coulomb potential $v(k)$ for this finite system.   Note that $\vwpol$ is automatically zero at $k=0$ as $\Wpol$ emanates from a polarization $\chi_\mathrm{pol}$ which vanishes at $k=0$ due to the orthogonality of the particle-hole pairs which make it. (For periodic systems this effect is counteracted by the singularity of the Coulomb potential at $k=0$, unlike finite systems, where $v(k=0$) is large but does not diverge so $\Wpol(k=0)$ vanishes.)  

In the bottom panel of Figure \ref{fig:vwk} we compare the ratio of $v_W(k)$ and $v(k)$.   The ratio is 1 for low $k$ due to the finite size of the systems, but levels down at higher $k$ values.


Table  \ref{tab:nv} shows the fraction of $\Wpol$ left for stochastic sampling after removal of $\vwpol$. This fraction is quite small and is clearly independent of system size. Additionally, it does not appear to change with the approximate dimensionality of the system-- linear, planar or spherical. Similarly, Figure \ref{fig:W_plot} shows, for a slice along the x-axis, the action of both $W$ and $v_W$ on two pair densities.  The results are very similar, but the total magnitude is often decreased when applying $v_W$.

We now turn to the spectra.  We used an iterative BSE Chebyshev procedure, and for all systems the upper bound on the half-width of the Liouvillian was taken as $\delta A = 16.5~$eV. We used $N_{\mathrm{cheby}}=500$ terms (Eq. \ref{eq:c_n_simple}), which is approximately equivalent to a Gaussian energy broadening with half width of 0.08 eV. The effect of the broadening is negligible for the larger systems where the spectrum is naturally quite broadened.

In Figure \ref{fig:spec} we show the spectra of all nine systems using a deterministic BSE,\cite{Bradbury2022} the $v_W$ only TD-OAI, and the stochastic $v_W + \{W-v_w\}$ approach of this paper. Using $v_W$ by itself is only qualitatively accurate, but stochastic sampling of $\{W-v_W\}$ quickly restores the accuracy of the deterministic calculation.

As is clear from Figure \ref{fig:conv}, at least $N_\beta=300-400$ stochastic samples are needed to get a  $0.1$~eV accuracy on the optical gap. Generally, the low-energies spectral peaks in Figure \ref{fig:spec} are converged to 0.02$~$eV at low energies by 2000 stochastic samples. (The one exception is fullerene, where the lowest-energy spectral peak converges to only $0.08~$eV at $N_\beta=5000$; this is in line with the lower quality of the $v_W$ fit to $W$ for fullerene, see Table \ref{tab:nv}.)  

The rapid convergence with $N_\beta$ implies that the stochastically sampled $\{W-v_W\}$ is generally numerically superior to the deterministic approach for systems with more than $\approx 70$ calculated valence orbitals.  This is because of the $N_v^2/2$ scaling of the number of pairs $W_{ij}$ when using directly the deterministic approach, Eq. (\ref{eq:bse_det}).

\begin{table*}[]
\centering
\caption{Gaps (eV) from stochastic DFT at the LDA level, stochastic $\mathrm{G_0W_0}$, self consistent $\mathrm{\Delta GW_0}$,\cite{Vlek20183} the stochastic BSE optical gap (this work), and a reference experimental optical gap. }
\label{tab:gap}
\begin{tabular}{|C{2.2cm} |C{1cm}C{1cm}C{1cm}C{1cm}|C{1.5cm}C{1.5cm}|}
\hline
    & DFT  & $\mathrm{G_0W_0}$ & $\mathrm{\Delta GW_0}$ &BSE  & \multicolumn{2}{|c|}{Experimental Optical Gap} \\
\hline
Nap                      & 3.4          & 7.6                             & 8.0          & 4.3          & 4.1                        & \cite{Costa2016, Menon2019}                \\
Tet                      & 1.6          & 5.1                             & 5.4          & 2.7          & 2.6                        & \cite{Costa2016, Menon2019}                \\
Hex                      & 0.8          & 3.7                             & 3.9          & 1.8          & 1.9                        & \cite{Tnshoff2010, Krger2017, Tnshoff2020} \\
Oct                      & 0.4          & 2.9                             & 3.1          & 1.3          & 1.5                        & \cite{Tnshoff2020, Krger2017, Mondal2009}  \\
Cor                      & 3.2          & 6.7                             & 7.1          & 4.3          & 3.7                        & \cite{Rouill2008}                          \\
$\mathrm{C_{60}}$        & 1.7          & 4.4                             & 4.7          & 2.3          & 1.8                        & \cite{Rabenau1993, Lof1995}                \\
Kek                      & 2.1          & 4.8                             & 5.1          & 3.2          &                            &                                                             \\
$\mathrm{C_{96}H_{24}}$  & 1.2          & 3.0                             & 3.1          & 1.9          & 2.0                        & \cite{Liu2022}                             \\
10-CPP+$\mathrm{C_{60}}$ & 0.7          & 3.3                             & 3.5          & 3.5          & 3.4  $^\dagger$                      & \cite{Xu2018}   \\
\hline
\end{tabular}
\newline
\footnotesize{ $\dagger$ Stabilized system complex.}

\end{table*}

In Table \ref{tab:gap} we summarize the evolution of the gap for each system.  The results are in fair agreement with the experimental values, considering the Tamm-Dancoff approximation and the lack of dynamic corrections.

\section{Conclusions\label{sec:con}}

We introduced here an optimized effective potential $v_W$ to reduce the magnitude of the $W$ term in BSE, enabling an efficient stochastic evaluation. With the introduction of $v_W$, the required number of stochastic orbitals is small relative to system size, thereby reducing the scaling of the method  so that large system sizes are now feasible. The new algorithm was checked successfully on nine molecules of varying dimension and size.

The present work overcomes the cost of the most expensive part in the BSE algorithm, preparing the action of $W$ on the product states, by dividing $W$ to an exchange-type potential $v_W$, and a stochastically sampled remainder $\{W-v_W\}$. There is, however, a lot of room for further scaling improvements. Currently, we do not implement the  exchange in a particularly efficient way, so that the scaling is still cuic, but there are many known techniques to dramatically improve this portion of the Hamiltonian, such as a fully stochastic exchange. \cite{Neuhauser2015sX,Xu2018,Romanova2022} Similarly, for both the exchange and the Coulombic part, i.e., the matrix elements in Eq. (\ref{eq:Afij}), a localized basis set would have reduced the scaling.  Once these two improvements are made the overall scaling of the method would reach quadratic.~\cite{Frster2021,Frster2022}

Further work on this method will include  fitting $W$ to give a $v_W$ interaction that goes beyond a translationally invariant interaction but preserves the quasi-linear scaling of $\int v_W(r,r')\beta (r') dr'$. An improved fit would make it possible to use very few stochastic samplings of the difference operator $\{W-v_W\}$ or just forego this term completely, keeping only $v_W$. 

 Further improvements include the anti-resonant to resonant transition couplings to go beyond the Tamm-Dancoff approximation.  As mentioned in Section \ref{sec:alg}, since the contribution of this  `off-diagonal' coupling in the BSE is substantially smaller than that of the resonant $W$, they could be represented by $v_W$ alone rather than  the full $W$, so no addtional $W$ samplings would be needed.

Lastly, dynamical corrections are needed in many systems with dominant $n\to\pi^*$ and $\pi\to\pi^*$ excitations.~\cite{Ma2009,Baumeier2012} While recent work has shown best results with a matrix perturbation theory based methods,\cite{Loos2020dyn} TDDFT-type approaches have been successful at capturing double excitations with a dynamical exchange kernel.~\cite{Romaniello2009, HuixRotllant2011,Rebolini2016} 

To summarize, the optimized attenuated potential reduces the magnitude of the effective interaction $W$.  This reduces the required number of stochastic sampling of $\{W-v_W\}$ to a manageable number, in the few thousands, enabling efficient BSE simulations. Further, when fitting $W$ to a translationally-invariant (convolution) interaction, the resulting TDHF spectra with $v_W$ as the exchange interaction are in quite good agreement with the exact $W$-based BSE results.

\section*{Acknowledgements}
We are grateful for discussions with Vojtech Vleck. This work is supported by the U.S. Department of Energy, Office of Science, Office of Advanced Scientific Computing Research, Scientific Discovery through Advanced Computing (SciDAC) program under Award Number DE-SC0022198. NCB acknowledges the National Science Foundation Graduate Research Fellowship Program under grant DGE-2034835.  Computational resources were provided by the National Energy Research Scientific Computing Center, a DOE Office of Science User Facility supported by the Office of Science of the U.S. Department of Energy under Contract No. DE-AC02-05CH11231 using NERSC award BES-ERCAP0020089.

\bibliography{main}
\end{document}